\documentclass[epsfig,floats,prl,twocolumn,showpacs,
superscriptaddress,preprintnumbers,floatfix]{revtex4-1}
\usepackage{graphicx}
\usepackage[latin2]{inputenc}
\usepackage{amsmath}
\usepackage{amssymb}
\usepackage{bm}
\usepackage{color}

\begin{document}

\title{Effect of plasticity and dimensionality on crumpling of a thin sheet}

\author{Mehdi Habibi}
\email{M.Habibi@uva.nl}
%\affiliation{Department of Physics, Institute for Advanced
%Studies in Basic Sciences, Zanjan 45195, Iran}
\affiliation{Van der Waals-Zeeman Institute, University of
Amsterdam, 1098 XH Amsterdam, The Netherlands}

\author{Mokhtar Adda-Bedia}
\affiliation{LPS, ENS, 24 Rue Lhomond, 75005 Paris, France}
\author{Daniel Bonn}
\affiliation{Van der Waals-Zeeman Institute, University of
Amsterdam, 1098 XH Amsterdam, The Netherlands}

\begin{abstract}
The process of crumpling a sheet and compacting it into a ball is dependent on many parameters that are difficult to disentangle. We study the effect of plasticity on the crumpling process, and disentangle the effects of plasticity and dimensionality of compaction by performing isotropic compaction experiments on various materials with different elastoplasticities. The force required to crumple a sheet into a ball as well as the number of layers inside the ball have a power-law dependence on the size of the crumpled ball, each with its own power-law exponent. We experimentally determine both exponents and find that they are linearly proportional to and decrease with increasing plasticity of the material. Finally we provide a scaling argument predicting a linear relation between the two exponents with a coefficient of 3.5 in excellent agreement with our experimental results.
\end{abstract}

\pacs{ 62.20.F-, 89.75.Da, 46.25.-y}

%62.20.F-   Deformation and plasticity
%89.75.Da   Systems obeying scaling laws
%46.25.-y   Static elasticity

\maketitle
The phenomenon of crumpling is not only experienced frequently in everyday life, it is also a ubiquitous phenomenon in nature and technology: from cortical folding in mammalian brains \cite{mota}, DNA packing in viral capsids \cite{DNA-viral}, flower buds \cite{Kobayashi98} to crumpled graphene \cite{graph}.
Crushing and squeezing a piece of paper into a ball results in a very light structure (with more than 80$\%$ void) with a complex fractal topology \cite{Balankin2010}, negative Poisson's ratio, surprising mechanical strength and the ability to absorb mechanical energy \cite{Deboeuf13,Balankin2015}. These peculiar mechanical properties make crumpled sheets a strong candidate for designing mechanical metamaterials. Furthermore, disordered crumpled structures -- in contrast to ordered origami structures -- benefit from inherent insensitivity to noise and defects which can result in superior mechanical functionality in real-world applications. However, key fundamental questions about crumpled structures need to be addressed before their potential can be fully exploited. In spite of extensive experimental \cite{Matan, Blair05,Lin09,Cambou11,Aharoni10,Balankin2007,Balankin}, theoretical \cite{Lobkovsky,Das,Kramer,Witten98,Sultan06,Adda-Bedia10,Benamar97} and numerical \cite{Vliegenthart06,Tallinen09,Ast04,Tallinen08} work devoted to study the crumpling mechanism, different aspects of this complex phenomenon are still elusive or controversial, and a general physical understanding is lacking. This is mainly due to the formation of complex random networks of localized stress \cite{Kramer,Witten98,Benamar97}, with highly nonlinear deformed regions influenced by plastic deformation, self-avoiding interactions and jamming effects \cite{Sultan06}, which make this phenomenon notoriously difficult to study.
It is known that the crumpling force increases as a power law with the compaction ratio, with the power-law exponent depending on the dimensionality of the compaction \cite{Matan, Lin09,Cambou11,Aharoni10,Deboeuf13} and self-avoiding interactions \cite{Vliegenthart06}.

Numerical studies to identify the contribution of self-avoidance to the
morphological properties of crumpled objects \cite{Vliegenthart06,
Tallinen09,Balankin,Sultan06} yielded that for 2D sheets the
mass-size relation is dependent on many parameters such as self-avoidance,
plasticity, and dimensionality of compaction
\cite{Vliegenthart06,Tallinen09, Sultan06} - parameters that are difficult
to disentangle. Indeed, elasto-plasticity is present in realistic materials, and localization of large stress and strain in very small regions leads to plastic deformation at the ridges. It has been suggested by \cite{Balankin2007} that the plasticity affects the crumpling only slightly. Numerical studies of crumpling predict that plasticity results in different morphologies and fractal dimensions when compared to purely elastic systems, but not to different power-law exponents: the crumpling force varies with the compaction ratio according to a power law with an exponent of -3.83 independent of the plasticity (for isotropic compaction in 3D) \cite{Tallinen09}.

It has been recently shown that the crumpling process can be viewed as arising from successive folding events and that predictions from simple folding model can capture complicated features of the crumpling process \cite{Deboeuf13}. Based on this model, crumpling exponents of -4 and -6 are predicted for 3D isotropic crumpling of elastic and plastic sheets, respectively. Here, the plasticity of the material \it{does} affect the crumpling process.

\begin{figure}[b]
\centering
\includegraphics[trim=0.0cm 0.9cm 0cm 1.2cm,width=\columnwidth]{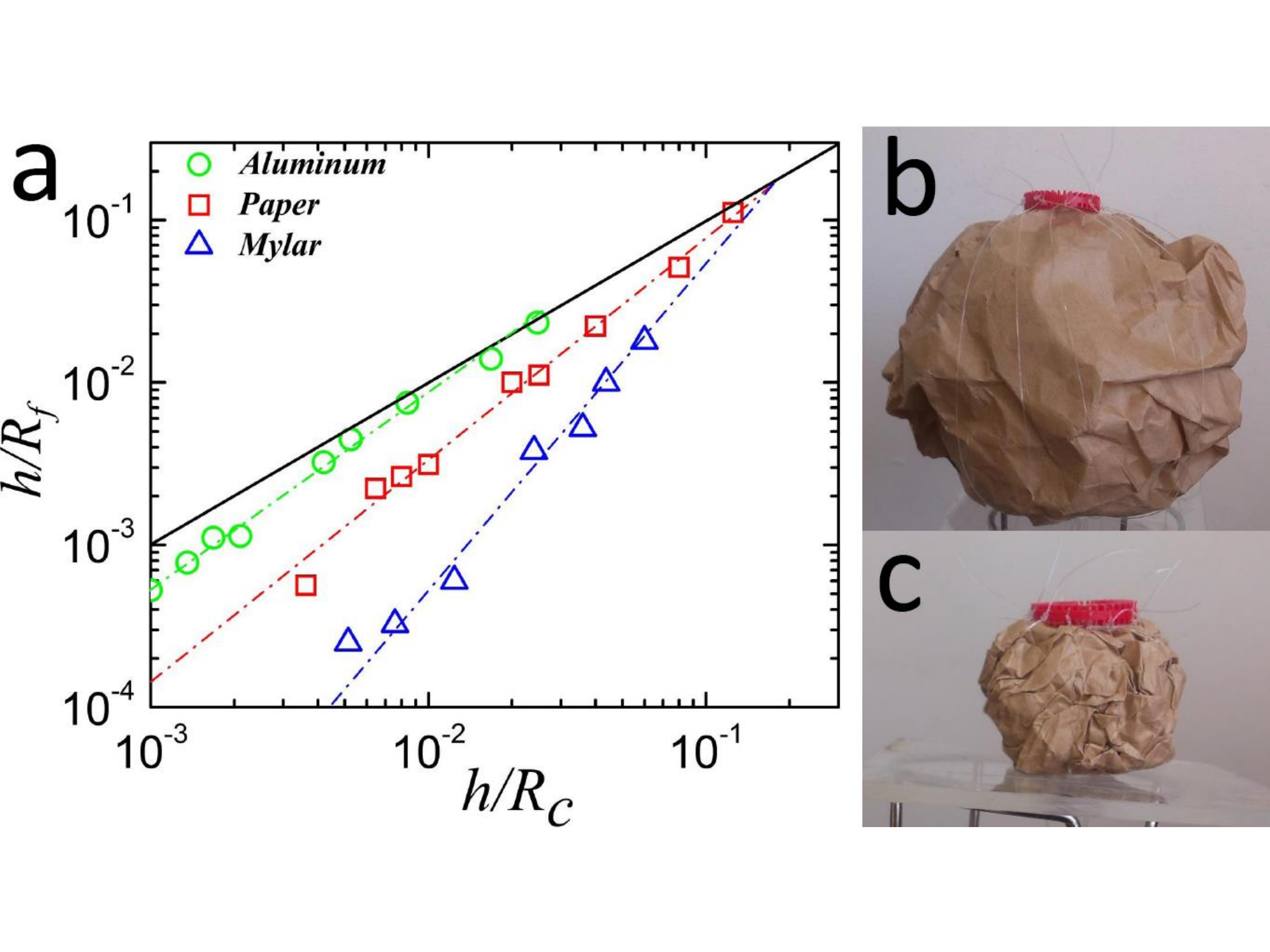}
\caption{a) Dimensionless curvature of a ribbon ($h/R_{f}$) as a function of dimensionless curvature of a cylinder ($h/R_{c}$) on a double logarithmic scale.  The solid line represents an ideal plastic material where $h/R_{c}\sim h/R_{f}$.  The inverse of the power-law exponent is considered as plasticity index of the material ($b$). b) Experimental setup consisting of a net of wires distributed uniformly around a loosely crumpled sheet, which pass through a hole in the platform on which the crumpled sheet is placed.  c) By hanging weights at the bottom of the net we arrive at a higher degree of compaction in 3D.}
\label{Fig1}
\end{figure}

In this work we aim to study the effect of plasticity on the crumpling process and disentangle the effects of plasticity on the one hand and dimensionality of compaction on the other, by performing isotropic compression of sheets of various materials with different elasto-plasticities in 3D. We present experimental measurements of the crumpling force as a function of size of the crumpled ball and study how the power-law crumpling-force exponent depends on the plasticity of the material. Finally, we study the power-law dependency of the number of layers on the size of a crumpled ball and show that the crumpling-force exponent is linearly proportional to the layering exponent, and that both exponents decrease with increasing plasticity.

{\it Experimental procedures}

{\it a)	Plasticity measurements---} Square sheets (with initial size $D_{0}=30~cm$) of different materials including PDMS rubber (with thickness $h=1~mm$), Mylar ($h= 75, 36, 23$ ~and ~19~$\mu m$), regular printing paper ($h=100~\mu m$) and aluminum foil ($h$=20 ~and ~8~$\mu m$) are used to cover a wide range of ductilities. To quantitatively measure the ductility of these materials we introduce a novel method, in which an initially flat thin ribbon of the material ($2~cm$ in width) is rolled around a cylinder of radius $R_{c}$ (ranging from $1$ to $50~mm$) under a constant extensional stress. The rolled ribbon is kept for about 30 minutes (comparable to maximum relaxation time of the materials used \cite{Lin09,Thiria}) under a constant tension and then released. After releasing the stress, the ribbon reaches a final radius of curvature $R_{f}$. $R_{f}$ is measured about 30 minutes after releasing the stress.  In Fig.~\ref{Fig1}, the dimensionless curvature of the ribbon ($h/R_{f}$) is plotted as a function of the dimensionless curvature of the cylinder ($h/R_{c}$) on a double logarithmic scale for different materials. $h/R_{f}$ shows a power-law dependence on $h/R_{c}$  with an exponent that increases with increasing ductility of the material. Here, the inverse of the power-law exponent is considered the plasticity index of the material $b$. For an ideal plastic material one expects $h/R_{c}\sim h/R_{f}$ which gives $b=1$. For an ideal elastic material the final curvature is always zero and independent of the curvature of the cylinder, therefore $b=0$. Fitting power laws to the experimental data in Fig.~\ref{Fig1} gives plasticity indexes of $0.86\pm 0.05$, $0.74\pm 0.05$ and $0.55\pm 0.1$ for the aluminum foil, paper, and Mylar, respectively.

{\it b)	Experimental setup---} The experimental setup consists of a net of wires distributed uniformly around a loosely crumpled sheet, which goes through a hole as shown in Fig.~\ref{setup}. The amount of confinement is increased by hanging weights at the bottom of the net. This setup enables us to crumple the sheet in an isotropic manner in 3D. About 10 minutes after adding weight to the net, the crumpled sheet reaches its approximate final size, which does not change considerably in time. Then the average crumpled size and the crumpling force are recorded, and extra weights are hanged on the net to arrive to a higher level of compaction. The contact points between the wires and the hole are lubricated using Silicone oil (100 cSt) to reduce friction effects.

\begin{figure}[b]
\centering
\includegraphics[trim=0.0cm 0.9cm 0cm 1.2cm,width=\columnwidth]{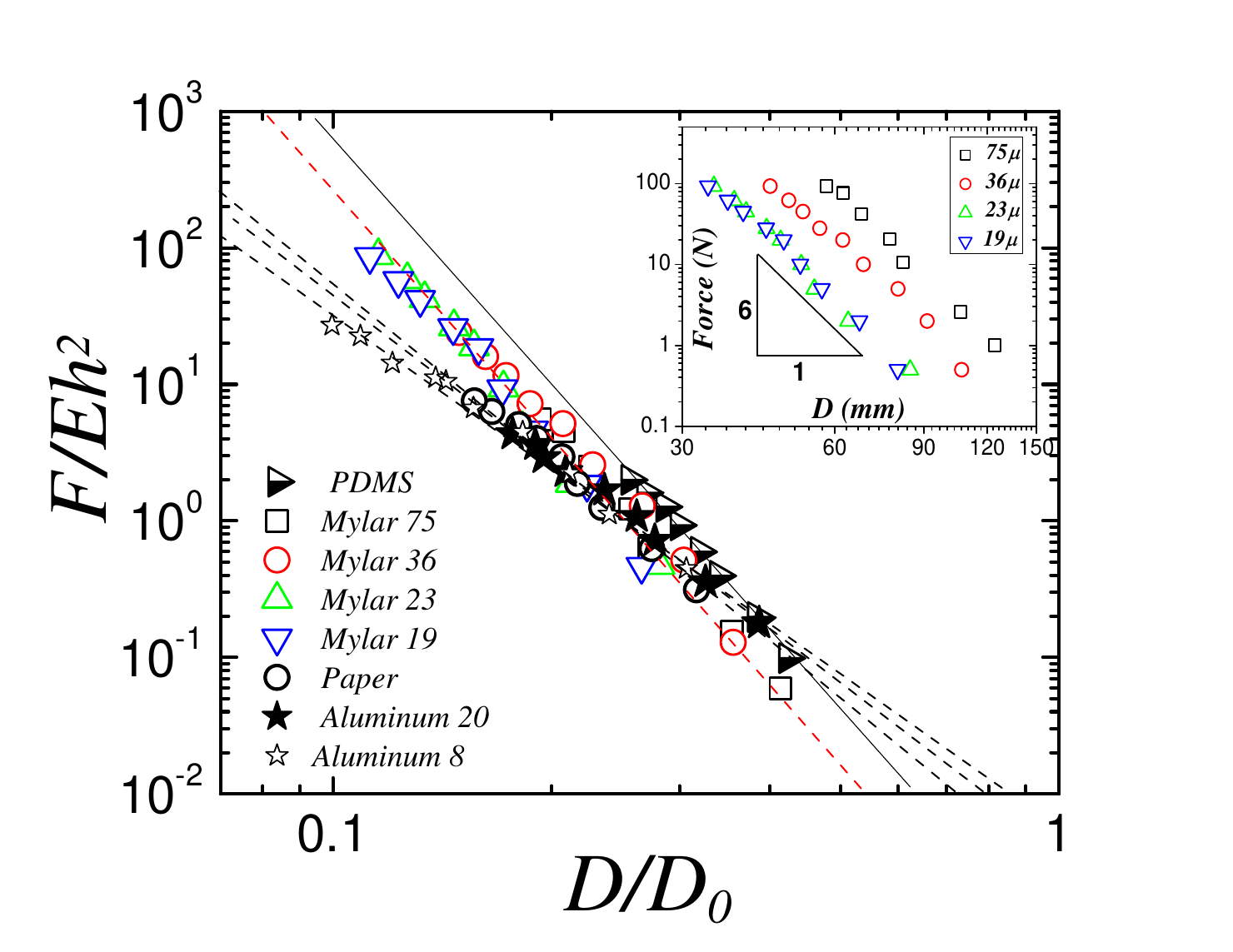}
\caption{Dimensionless crumpling force ($F/Eh^{2}$) as a function of dimensionless crumpled size ($D/D_{0}$). The size-dependency of the rescaled crumpling force is captured well by $F/(Eh^{2})\sim(D/D_{0} )^{-\beta}$ with a plasticity-dependent exponent $\beta$. Inset shows crumpling force $F$ as a function of the average crumpled size $D$ for Mylar sheets of different thicknesses.}
\label{Fig2}
\end{figure}

{\it Experimental results}

{\it a)	Plasticity-dependent exponent---} In the inset of Fig.~\ref{Fig2} the crumpling force $F$ is plotted as a function of the average crumpled size $D$ for Mylar sheets of different thicknesses. As expected, $F$ increases with decreasing the size of crumpled ball in a power-law fashion. The crumpling-force exponent for Mylar is about -6 for different thicknesses. The observed shift for different series of data in Fig.~\ref{Fig2} is due to differences in the thickness of the sheets and indicates that the crumpling force can be rescaled using a scaling relation for the force needed to make a single fold ($Eh^{2}$, where $E$ is the elastic modulus and $h$ the thickness of the sheet) \cite{Deboeuf13}. Rescaled data for all the materials are shown in Fig.~\ref{Fig2}, where data points for different materials collapse onto each other.  The size-dependency of the rescaled force is captured well by $F/(Eh^{2} )\sim(D/D_{0} )^{-\beta}$ with a plasticity-dependent exponent $\beta$. This exponent (obtained by fitting power laws to the experimental data) decreases with increasing plasticity index ($b$) from 6 for the PDMS sheet to 3.9 for the aluminum foil, as shown in Fig.~\ref{Fig4}. The exponent of 3.9 for aluminium foil is in good agreement with the numerical prediction of \cite{Tallinen09}. In reference \cite{Lin09}, 3D compaction of a sheet under ambient pressure was studied and $\beta$ exponents of 5.1 and 15.4 were reported for aluminium foil and HDPE polymer sheets, respectively. To compare our results with \cite{Lin09} we have performed the same compaction procedure under ambient pressure for aluminium foil and Mylar, and obtained $\beta$ exponents of 5.4 $\pm$ 0.5 and 16 $\pm$ 1, respectively (data not shown here). However, we believe that the $\beta$ exponents obtained with this method are not reliable because the crumpled structures are not isotropic especially for elastic sheets like Mylar and HDPE. In addition, due to experimental limitations  large areas of the sheet have to be crumpled by hand before starting the experiment   which can considerably affect the measured $\beta$.
As mentioned in the introduction, the folding model predicts values for $\beta$ of 4 and 6 for ideal elastic and ideal plastic materials, respectively, in contrast with our experimental observation \cite{Deboeuf13}.

\begin{figure}[b]
\centering
\includegraphics[trim=0.0cm 0.9cm 0cm 1.2cm,width=\columnwidth]{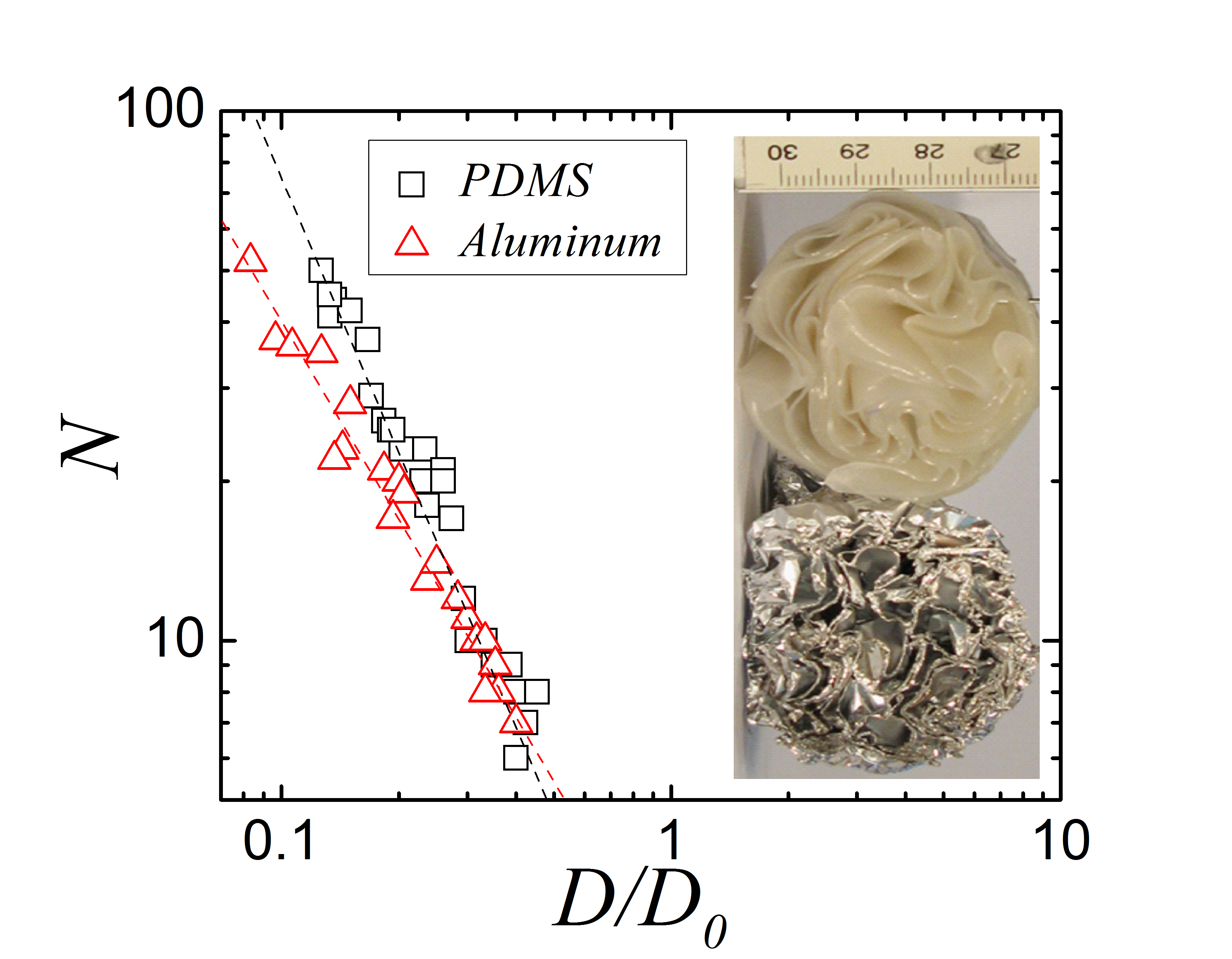}
\caption{Number of layers $N$ as a function of the dimensionless average size of the crumpled ball ($D/D_{0}$) for PDMS rubber and aluminum foil. Data points for Mylar and paper are not shown on the graph for sake of clarity. The inset shows cross sections
of crumpled sheets of aluminum foil and rubber; for aluminum foil the outer layers are more compact and have smaller radius of curvature. For rubber the compaction is more homogenous and the average radius of curvature is larger. }
\label{Fig3}
\end{figure}

{\it b)	Morphology and number of layers---} A second step towards understanding the crumpling exponent is to establish a relation between the degree of compaction and the number of layers for the crumpled configurations. To achieve this, sheets of different materials are crashed into hand-made balls at different degrees of compaction. The number of layers is measured either from layer counting on the cross-section (by cutting the crumpled ball in two) or by passing a needle through the crumpled ball and counting the number of holes in the sheet after un-folding it. The number of layers is measured in three orthogonal directions passing through the center and averaged to obtain the mean number of folded layers ($N$).  Fig.~\ref{Fig3} interestingly reveals a power-law dependency of $N$ on the crumpled ball size, of the form  $N\sim (D/D_{0})^{-\gamma}$. The exponent $\gamma$ obtained by fitting the experimental data changes slightly with changing ductility of the material and decreases from $1.7\pm0.1$ for the rubber to $1.5\pm0.1$  for the paper and finally reaches to $1.2\pm0.1$ for the aluminum foil (for sake of clarity data for the Mylar and paper are not shown in Fig.~\ref{Fig3}). It worth mentioning that the folding model of Deboeuf {\it et al.} \cite{Deboeuf13} predicts an exponent of 2 for ordered folding in 3D.

\begin{figure}[b]
\centering
\includegraphics[trim=0.0cm 0.9cm 0cm 1.2cm,width=\columnwidth]{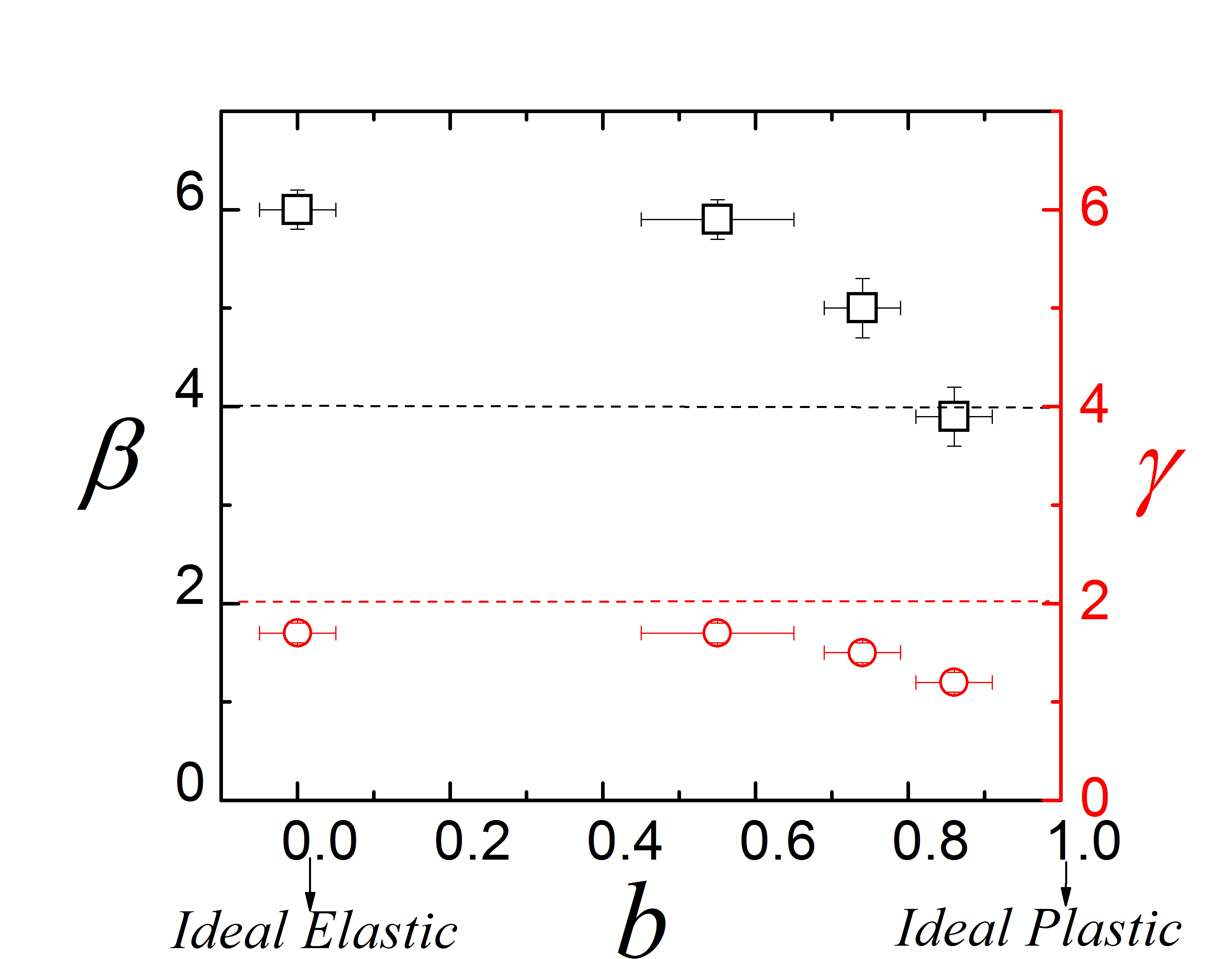}
\caption{Crumpling exponent $\beta$ and layering exponent $\gamma$ as a function of the plasticity index $b$. Both exponents are obtained by fitting power laws to the experimental data for different materials. The plasticity index is 0 (1) for an ideal elastic (plastic) material.  }
\label{Fig4}
\end{figure}

\begin{figure}[b]
\centering
\includegraphics[trim=0.0cm 0.9cm 0cm 1.2cm,width=\columnwidth]{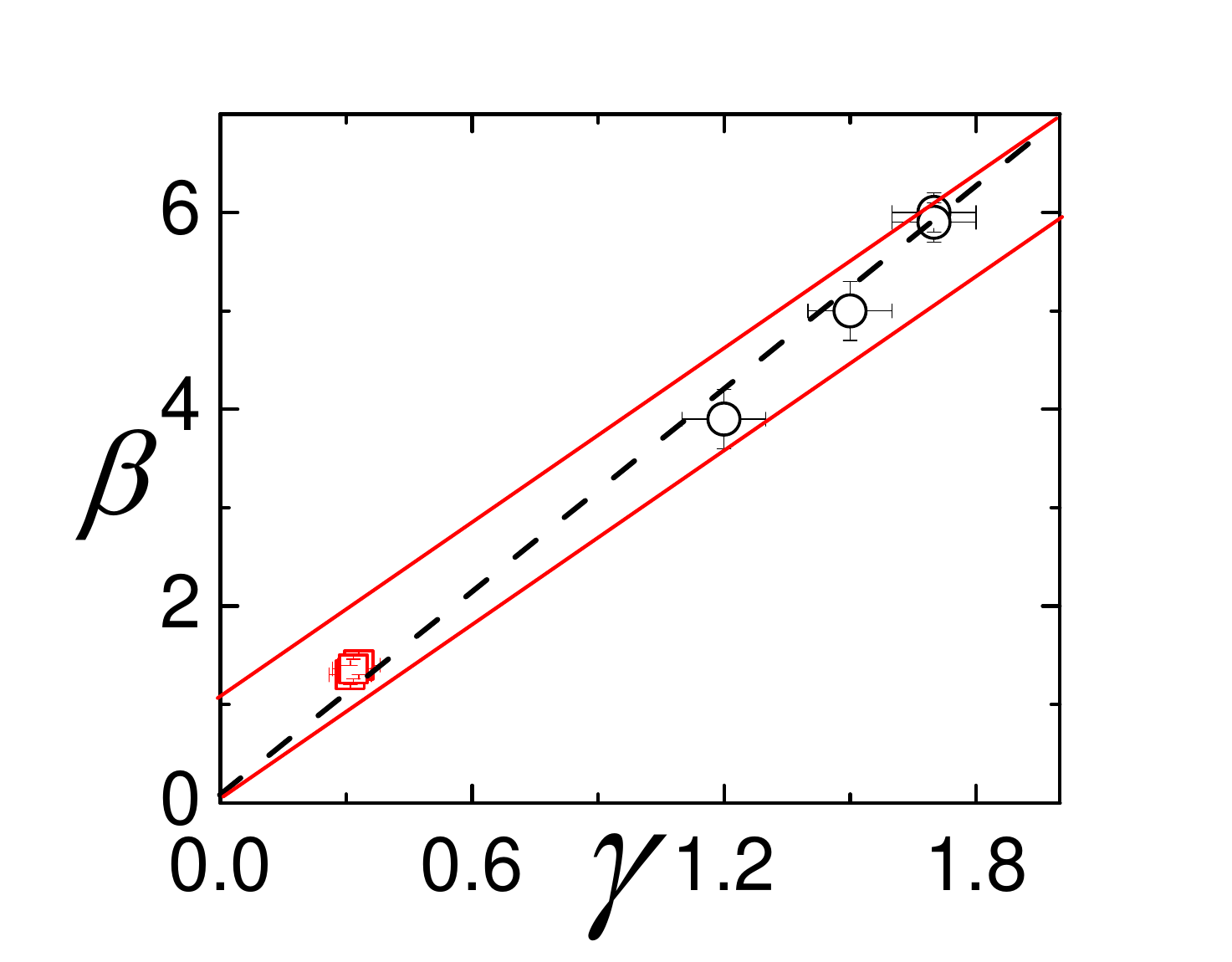}
\caption{Crumpling exponent $\beta$ as a function of layering exponent $\gamma$ for different materials.  The dashed line with slope 3.47 is the best fit to the data when it is forced to go through the plot's origin. Red lines indicate the upper and lower limits predicted by the scaling.}
\label{Fig5}
\end{figure}

Plotting the experimental values for $\beta$ as a function of $\gamma$ for different materials (Fig.~\ref{Fig5}) reveals a linear dependency. In this plot circles are related to isotropic 3D compaction of PDMS, Maylar, paper and aluminium foil while squares represent experimental results for quasi 1D compaction of Maylar, paper and aluminium foil in a cylinder. The experimental details for quasi 1D compaction are the same as described in \cite{Deboeuf13}. The best fit to all the data (assuming a linear dependency with no offset) has an slope of 3.47.

{\it Discussion---} By crushing a crumpled ball of elastic material there will be a disordered network of creases and folds. Elasticity theory can be used to predict a relation between the compaction force and crumpled size by assuming that every piece of the sheet in the crumpled ball can be buckled under the compaction force. According to Euler's buckling theory, the force needed for buckling a plate scales as $(EH^{3})/R$ where $R$ is the size and $H$ is the thickness of the plate \cite{Euler,Deboeuf13}.  In a crumpled ball, the maximum size of the plate is set by the size of the crumpled ball ($D$) while the minimum bending radius has the same order of magnitude as the thickness of the sheet $h$ in a single fold. Using the power-law relations for $F$ and $N$ as a function of $D$, and by replacing $H$ with $Nh$, and $R$ with $D$ ($h$) in the above equation one arrives at $\beta=3\gamma+1$ ($\beta=3\gamma$) as upper (lower) limit for the force exponent.  These predictions are shown by two red lines on Fig.~\ref{Fig5}. All the experimental data indeed lie between the two limits. A more realistic estimation for the size in a crumpled ball is the average size of creases ($\bar{l}$) which is  proportional to $N^{-0.5}$, because $N\bar{l}^{2}$ is proportional to the total area of the sheet (which is constant in our experiments). Thereby replacing $R$ with $\bar{l}$ in the previous relation results in $\beta=3.5\gamma$, which is in excellent agreement with the experimental results shown in Fig.~\ref{Fig5}. A similar linear relation between force exponent and layering exponent was predicted by the folding model of Deboeuf {\it et al.} \cite{Deboeuf13} with coefficients of  2, 1, and 1 for 3D, 2D and 1D folding, respectively. These findings indicate that the proportionality coefficient between compaction force exponent and layering exponent is independent of the material properties but set by the method of compaction (crumpling or folding).

Furthermore, plasticity can affect the homogeneity of the radius of the curvature of the layers as well as the number of layers. High plasticity of the material results in the formation of irreversible creases. Consequently, the crumpling process results in some wrinkles with very small radius of curvature.  For an elastic sheet, these small-curvature creases will disappear when higher compactions are achieved, as shown in \cite{Poch}. Formation of irreversible wrinkles in plastic sheets can cause shrinkage of the effective area of the sheet. This consequently leads to a smaller number of layers at a certain degree of compaction -- in agreement with our experimental observations in Fig.~\ref{Fig3}.
We can also appreciate this effect from the cross sections of crumpled sheets of aluminum foil and rubber (inset of Fig.~\ref{Fig4}): for aluminum foil the outer layers are more compact and have smaller radius of curvature; for the rubber with the same degree of compaction the compaction is more homogenous and the average radius of curvature is larger.

In conclusion, we investigated crumpling of sheets of different materials under isotropic compaction and studied the effect of ductility of the material on the compaction mechanism.  We used a new method to measure ductility of a thin sheet by introducing a plasticity index based on irreversible deformation of the sheet. This definition and method enabled us to quantitatively determine the plasticity index for a series of materials from elastic (rubber) to very plastic (aluminum foil). We measured the force needed to crush thin sheets of different materials into crumpled balls as well as the average number of layers in the crumpled structures as a function of the diameter of crumpled ball. Both quantities were shown to increase with decreasing cumpled-ball diameter according to power laws, with a plasticity-dependent exponent for each quantity. We found that both exponents are linearly proportional to and decrease with increasing plasticity of the material. Finally, based on elasticity theory, we developed a scaling argument predicting a linear relation between the two exponents with a coefficient of 3.5 in excellent agreement with our experimental results.

The results provide a new insight into underlying mechanisms of crumpling and will be of wide interest to researchers in the field of biophysics, nanotechnology ({\it e. g.} crumpled graphene) and mechanical metamaterials.

We gratefully acknowledge A. D. Cambou and N. Menon for helpful discussions. This work is part of the research program of FOM, which is financially supported by NWO.

\end{document}